\documentstyle[aps,prl,preprint]{revtex}
\draft

\begin{document}
\title{New valence bond crystal phase of a frustrated spin-$\case{1}{2}$  
square lattice antiferromagnet}
\author{M. E. Zhitomirsky$^{1,2}$ and Kazuo Ueda$^1$}
\address{
$^1$Institute for Solid State Physics, University of Tokyo,
Tokyo 106, Japan  \\
$^2$Landau Institute for Theoretical Physics, Moscow 117334, Russia}
\date{March 15, 1996}
\maketitle

\begin{abstract}
We propose a new type of magnetically disordered ground state
for a frustrated quantum antiferromagnet. This 
disordered state is an array of spin singlets
spontaneously formed on four spin plaquettes. 
Both perturbation results and bond-operator calculations
show that this phase has lower
energy than the columnar dimer state. Analysis of available
numerical data on finite clusters also supports the conclusion
that this state is realized at intermediate frustrations.
\end{abstract}

\narrowtext

Frustrated two-dimensional quantum magnets are a fascinating 
topic of numerous studies over the past decade. The interest
has primarily focused on melting of long-range magnetic order and     
appearance of disordered phases at $T=0$ as a result of enhanced 
quantum fluctuations. The most popular spin system having 
such type of behavior is a frustrated square lattice 
Heisenberg model with the nearest-neighbor antiferromagnetic
exchange $J_1$ and the second-neighbor coupling $J_2$
\begin{equation} 
{\cal H} = J_1 \sum_{\rm NN} {\bf S}_i\cdot {\bf S}_j
    + J_2 \sum_{\rm 2NN} {\bf S}_i \cdot {\bf S}_j \ .
\label{H}
\end{equation}
Analysis within the linear spin-wave theory reveals that for
weak frustration the model has a N\'eel ground state with 
ordering wave-vector $(\pi,\pi)$, while for strong frustration
spins are ordered at wave-vector $(\pi,0)$ or 
$(0,\pi)$ (stripe or collinear state). 
For any finite $S$ there is also a region 
around the classical critical point $J_2/J_1=0.5$ 
where sublattice magnetization 
vanishes \cite{LSWT}. This result raises a question about the ground 
state for intermediate frustrations which may be nonmagnetic
and represent a kind of two-dimensional
spin liquid.

Subsequent theoretical works have considered the problem of 
the intermediate phase of the frustrated antiferromagnet
Eq.~(\ref{H}) mainly
from three different points of view.
The first group includes exact diagonalization studies
on finite clusters, which clearly show existence of a disordered
spin state in the region $0.4<J_2/J_1<0.65$, 
though final conclusion about its nature has not been
reached \cite{exact1,exact2}. 
The second group of works address the problem
by calculating higher-order $1/S$ corrections either in 
framework of the modified spin-wave theory or using the Schwinger
boson mean-field calculations \cite{MFSWT,Schwb}. Both approaches predict
enhanced stabilities for the N\'eel and the collinear states
resulting in a finite
overlap of the magnetically ordered phases.
Second-order corrections to the mean-field solution,
on the other hand, suggest
a small window $0.52<J_2/J_1<0.57$ between the two ordered 
states for $S=\case{1}{2}$ \cite{Gochev}. 

The third group of works explore one particular possibility
for disordered ground state---a valence bond crystal
with broken translational symmetry and a finite gap in 
excitation spectrum. Studying nearest-neighbor SU$(N)$
antiferromagnet in the large $N$ limit, Read and Sachdev have
argued in favor of a state with spontaneous 
columnar dimerization \cite{SUN}.
Subsequent calculations for the spin-$\case{1}{2}$ Hamiltonian (\ref{H}) 
using series expansion \cite{series} 
and boson techniques \cite{slbs} supported 
stability of this phase 
around $J_2/J_1=0.5$.
Consistent with these treatments cluster  
results for dimerized
susceptibility yield its noticeable increase
in the intermediate region \cite{exact1,exact2}.

In this Letter we propose a new type of valence bond crystal
ground state for a frustrated quantum 
antiferromagnet. This is a so called plaquette
resonating-valence-bond (RVB) state shown in Fig.~1(a).
We calculate its energy by two methods and find that
it is lower than for the columnar dimerization [Fig.~1(b)] in
the same approximation.
Then, by considering relevant order parameters,
we show that the existing numerical data on the $6\times6$ cluster
\cite{exact2} are, as a matter of fact, 
in favor of this state.

Two types of magnetically disordered singlet ground
states have been widely discussed for a spin-$\case{1}{2}$
square lattice antiferromagnet.
The first, a featureless RVB spin-liquid, 
describes a linear superposition of spin pair 
singlets with long- or short-range correlations \cite{RVB}. 
The second proposal
is a spin-Pierls order of valence bonds which are frozen and 
break lattice symmetries \cite{SUN}.
One can also consider
an intermediate situation: valence bonds resonate in finite
spin blocks resulting again in breaking of translational symmetry.
In the simplest case such spin blocks contain four sites.
Similarly to discussion of dimerized phases, the first step 
is to investigate  a restricted Hamiltonian of four spins on a
square plaquette:
\begin{equation} 
  {\cal H}_{\rm plq}\!=\!J_1 ({\bf S}_1\!+\!{\bf S}_3) \! \cdot \! 
   ({\bf S}_2 \! + \! {\bf S}_4)
   +  J_2 ({\bf S}_1 \! \cdot {\bf S}_3 \! 
+ \! {\bf S}_2\! \cdot {\bf S}_4) .
\label{Hplq}
\end{equation}
This Hamiltonian is easily diagonalized. The ground level for
$J_2<J_1$ is a singlet 
characterized by the quantum numbers $S_{13}=S_{24}=1$
with the energy $E_s=-2J_1+\case{1}{2}J_2$.
Its wave function 
is  $|s\rangle=\case{1}{\sqrt{3}}(\{1,2\}\{4,3\} + \{1,4\}\{2,3\})$, 
where curly brackets denote singlet bond of a 
spin pair. This state can be considered as a RVB-like
state in a four spin subsystem with energy
lower than for a frozen dimer configuration, e.g., $\{1,2\}\{4,3\}$. 
Excited levels are three
triplets $|t_\alpha\rangle$, $|p_\alpha\rangle$, $|q_\alpha\rangle$
($\alpha=x,y,z$), $E_t=-J_1+\case{1}{2}J_2$, $E_p=E_q=-\case{1}{2}J_2$,
a quintet $|d_\nu\rangle$ ($\nu=1,...,5$), 
$E_d=J_1+\case{1}{2}J_2$, and another singlet
$|s'\rangle=\{1,2\}\{4,3\} -\{1,4\}\{2,3\}$, $E_{s'}=-\case{3}{2}J_2$, 
which crosses with $|s\rangle$ and becomes the ground state for $J_2>J_1$
(we will not be concerned with this case in our paper).

Four spin singlets form a plaquette 
covering of a two-dimensional lattice.
Previously, the plaquette-RVB state has been predicted for the 
disordered two-dimensional magnet CaV$_4$O$_9$, which is
described by a spin-$\case{1}{2}$ Heisenberg model on a 1/5-depleted
square lattice \cite{CaVO}. In that case singlets correspond to
spin blocks chosen by the lattice geometry. For the
translationally invariant Hamiltonian (\ref{H}) periodic
array of plaquette singlets appears spontaneously.
The ground state has in this case four-fold degeneracy
determined by broken translations along two sides and diagonal
of an elementary square.
Using simple rules for products of
dimer coverings \cite{RVB} one can show
that the overlap between the two states in Fig.~1 formed by
pure singlets decreases to zero as $(\sqrt{3}/2)^{N/4}$ with
increasing number of spins $N$. Consequently, 
the plaquette-RVB state cannot be represented as 
a superposition of two column states rotated by $90^\circ$
with respect to each other. Later, we 
consider spin order parameters, which have different
values for the two disordered phases.

Since the Hamiltonian (\ref{H}) includes interaction of spins
from different blocks, the ground state does not coincide with a simple 
product of block singlets
and nonzero expectation values $\langle{\bf S}_i\cdot{\bf S}_j\rangle$
appear between all nearest
neighbor pairs. To compare energies of the columnar dimer and
the plaquette-RVB phases we use 
a boson technique suited for perturbative analytical
expansion around the local spin singlets. It was used previously
to study dimerized phases  
\cite{slbs}. We first generalize this method 
by deriving simultaneously boson representations for 
dimer and plaquette spins and then calculate 
energies of the two states in the mean-field approximation.

Let us consider an arbitrary spin block with singlet
ground state and excited levels denoted by $|\mu\rangle$.
A spin ${\bf S}_n$ in a block is expressed in terms of the basis
states by
$$
{\bf S}_n = \langle\mu| {\bf S}_n |\nu\rangle\,Z^{\mu\nu}\ , 
$$
where $Z^{\mu\nu}$ is the projection operator
$|\mu\rangle\langle\nu|$, $n$ is a local spin index inside the block, and
the global block index in the lattice $i$ is omitted for simplicity. 
We derive first the matrix elements
in a subspace of the ground state singlet and the lowest
triplet states. These states form a complete set for dimers,
whereas for plaquettes this is only a part of the local basis. Later,
we partially take into account higher 
energy states of plaquettes. Rotational
invariance in spin space and time-reversal symmetry give
\begin{eqnarray*}
 & & \langle s|{\bf S}_n^\alpha|s\rangle = 0 \ ,  \ \ 
\langle s|{\bf S}_n^\alpha|t_\beta\rangle = 
\delta_{\alpha\beta} A_{st}^n \ , \
\nonumber\\ 
 & & \langle t_\alpha|{\bf S}_n^\beta|t_\gamma\rangle = 
i e^{\alpha\beta\gamma} A_{tt}^n \ , 
\end{eqnarray*}
where $e^{\alpha\beta\gamma}$ is the totally antisymmetric tensor
and $A_{st}^n$, $A_{tt}^n$ are real constants.
Using the explicit forms of singlet
and triplet wave functions one straightforwardly gets 
$A_{st}^n = (-1)^n/2,\ A_{tt}^n = 1/2$ for dimers,
and $A_{st}^n = (-1)^n/\sqrt{6},\ A_{tt}^n = 1/4$ for plaquettes.
Let us now define the vacuum $|0\rangle$ and four boson 
operators which yield the four physical states by
$|s\rangle=s^+|0\rangle$,  
$|t_\alpha\rangle=t_\alpha^+|0\rangle$.
The projection operators are,  then, expressed as
$Z^{st_\alpha} = s^+t_\alpha$,   
$Z^{t_\alpha t_\beta} = t_\alpha^+t_\beta$, and so on.
Block spins represented via these boson operators are 
\begin{eqnarray}
& & S_n^\alpha = \case{(-1)^n}{2}(s^+t_\alpha+t_\alpha^+s)
 - \case{i}{2} e^{\alpha\beta\gamma}t_\beta^+t_\gamma ,  
\ \ \text{for dimers}, 
\nonumber\\[-2mm]
\label{repres}
\\[-2mm]
& & S_n^\alpha = \case{(-1)^n}{\sqrt{6}}(s^+t_\alpha+t_\alpha^+s)
 - \case{i}{4} e^{\alpha\beta\gamma}t_\beta^+t_\gamma ,  
\ \ \text{for plaquettes}. \nonumber
\end{eqnarray}
Spin commutation relations are satisfied in the chosen subspace 
as long as the bosonic representation preserves the algebra of 
the projection operators:
$Z^{\mu\nu}Z^{\mu'\nu'} =
\delta_{\mu'\nu} Z^{\mu\nu'}$.
This requirement restricts the boson occupation numbers:
\begin{equation}
s^+s + \sum_\alpha t_{\alpha}^+t_{\alpha}^{} = 1 \ .
\label{constraint}
\end{equation}
With this constraint 
the Hamiltonian of a single block, a dimer or a plaquette,  
takes a form,
${\cal H}_B = E_s s^+s + E_t t_\alpha^+t_\alpha^{}$.
The simplest way to deal with Eq.~(\ref{constraint}) is to
impose it as `a constraint in average' enforced by  
a chemical potential. 

Further calculations follow closely the work by Sachdev 
and Bhatt \cite{slbs}. Using Eq.~(\ref{repres}) we rewrite
the Hamiltonian (\ref{H}) in terms of bond operators
for each of the two states and assume site independent chemical 
potential $\mu$ and condensate of singlets $\langle s_i\rangle 
= \bar{s}$. Interaction terms in the boson Hamiltonian
are classified by the number of triplet operators 
$t_{\alpha{\bf k}}$. It can be shown that the terms with three
and four triplet operators affect the results only slightly 
\cite{slbs}. Therefore, we omit them and after diagonalization  
of the remaining quadratic form 
determine the parameters $\mu$ and $\bar{s}$ 
through the saddle-point equations
\begin{equation}
\partial E_{\rm g.s.}/\partial\mu = 0 \ , \ \ 
\partial E_{\rm g.s.}/\partial\bar{s} = 0 \ . 
\label{saddle}
\end{equation}
The system is a magnetically disordered phase as long as the gap 
in the excitation spectrum is positive. Vanishing of the gap
leads to a condensation of triplet excitations:
$\langle t_{i\alpha}\rangle\neq 0$. Note, that this 
does not mean transition, e.g., to the usual N\'eel 
state, since spontaneous dimerization of bonds will be 
preserved in such an ordered phase.

For the columnar dimer state we recover the results of 
\cite{slbs}. This state is stable in the region
$0.19<J_2/J_1<0.66$. For the plaquette-RVB phase 
analogous treatment would underestimate the energy because of 
the neglection of the higher-lying states of plaquette.
Among the other excited levels the most significant energy corrections
are given by higher triplets which have nonzero 
matrix elements with the ground state for spin operators.
Extending calculations preceding Eq.~({\ref{repres}) to these
excited triplets we find the following 
representation of spins in each plaquette 
\begin{eqnarray}
& & S_{1,3}^\alpha = \case{1}{\sqrt{6}}(s^+t_\alpha+t_\alpha^+s) 
 \pm \case{1}{2\sqrt{3}}(s^+p_\alpha + p_\alpha^+s) \ ,
\nonumber\\[-2mm]
\label{represext}
\\[-2mm]
& & S_{2,4}^\alpha = -\case{1}{\sqrt{6}}(s^+t_\alpha+t_\alpha^+s) 
 \pm \case{1}{2\sqrt{3}}(s^+q_\alpha+q_\alpha^+s) \ ,
\nonumber
\end{eqnarray}
with the constraint (\ref{constraint}) being changed to
$s^+s + \sum_\alpha (t^+_\alpha t_\alpha + p^+_\alpha p_\alpha
+ q^+_\alpha q_\alpha)=1$. 
According to the quadratic approximation we have omitted
in Eq.~(\ref{represext}) products of two triplet operators.
The boson Hamiltonian obtained with this
substitution can be further simplified by keeping 
interaction terms with only one of the higher triplets 
$p_\alpha$ and $q_\alpha$. Physically, it means
that we take account of higher-lying 
triplets only via scattering of lowest excitations on them.
As a result, one mode in the upper band remains constant,
while the other acquires a finite dispersion.

Solution of the saddle-point equations (\ref{saddle}) gives us
parameters $\mu$ and $\bar{s}$, from which we calculate
energy and spin gap of the plaquette-RVB phase.
This disordered state is locally stable for $J_2>0.08J_1$.
For large frustrations stability boundary of this phase lies at 
$J_2>0.8J_1$, well inside the region where 
the collinear state is expected to appear. This boundary 
cannot be determined correctly in our approximation as 
it depends crucially on the behavior of the neglected 
higher singlet. Therefore, we restrict ourselves
to the region $J_2\le0.7J_1$.
The gap of the plaquette state
at $J_2=0.5J_1$ is $\Delta_{\rm plq}=0.85 J_1$, while 
for the columnar dimer state $\Delta_{\rm dmr}=0.74 J_1$. 
The results for the energies of these two phases 
in units of $J_1$ are presented in
Fig.~2 by solid lines. In addition to significantly
wider region of stability, the plaquette-RVB phase 
has lower energy than the dimerized
state for $J_2<0.58J_1$. At $J_2=0.5J_1$ the difference between the
two energies, $E_{\rm plq}=-0.466$ and $E_{\rm dmr}=-0.456$, 
is about 2\%. (Without higher triplets 
$E_{\rm plq}=-0.458$.) Thus, the bond-operator formalism predicts
the plaquette-RVB state instead of the columnar dimer state
as the intermediate phase of the frustrated antiferromagnet (\ref{H}).

Another way to estimate analytically the energies of the two valence 
bond states is the second order perturbation expansion 
starting from the singlets either on dimers
or on plaquettes. 
This approximation
corresponds to the first two nonzero terms in the series expansion
method \cite{series}. The results are shown in Fig.~2 by dashed lines. 
At $J_2=0.5J_1$ the difference between $E_{\rm plq}=-0.63$ 
and $E_{\rm dmr}=-0.492$ is much bigger than in the bond-operator 
scheme. Though significant corrections are expected in
higher orders of series expansions, 
these calculations agree with our proposal of 
the plaquette-RVB state for the ground state of 
the Hamiltonian (\ref{H}). 
They also show that the crossing of the two valence bond levels
found by the boson technique may be shifted to stronger
frustrations.

Both results for the energies of the two disordered phases have been
obtained by approximate analytical methods and, hence, may be questioned.
We, therefore, now discuss how the columnar and plaquette
phases can be distinguished in exact numerical diagonalization studies.
The key step is to construct appropriate
order parameters, which measure nonequivalence of spin bonds
in a disordered state.
A set of such spin operators have been 
proposed by Sachdev \cite{OP}. Appearance of  
the column phase has been tested  \cite{exact1,exact2}
by using the parameter
\begin{equation}
\Psi_{\bf i} \! = \! {\bf S}_{\bf i}\! \cdot \! [  
(\! -1\! )^{i_x}\!  ({\bf S}_{{\bf i}-\hat{\bf x}}\! - \! 
{\bf S}_{{\bf i}+\hat{\bf x}}) \! 
+ \! i (\! -1\! )^{i_y}\! (
{\bf S}_{{\bf i}-\hat{\bf y}} \! - \! 
{\bf S}_{{\bf i}+\hat{\bf y}})]\,.
\label{OP1}
\end{equation}
This operator is site independent and  
takes values proportional to 
$1$, $i$, $-1$, and $-i$ for the four 
degenerate column states. On the other hand, 
for the four plaquette-RVB states 
$\Psi_{\bf i}$ equals 
$e^{i\pi/4}$, $e^{i3\pi/4}$, $-e^{i\pi/4}$, and 
$-e^{i3\pi/4}$ up to a constant factor. 
Since in numerical studies on finite systems one measures only
the corresponding correlation function 
$\chi_\psi = 1/N^2\langle|\sum_{\bf i}
\Psi_{\bf i}|^2\rangle$, it is impossible to distinguish the
two states by observing an anomaly in $\chi_\psi(J_2)$.  
To overcome this difficulty 
another order parameter should be considered 
\begin{equation}
\Phi_{\bf i} = 
{\bf S}_{\bf i} \cdot ({\bf S}_{{\bf i}+\hat{\bf y}} +
{\bf S}_{{\bf i}-\hat{\bf y}}
- {\bf S}_{{\bf i}+\hat{\bf x}} - 
{\bf S}_{{\bf i}-\hat{\bf x}}) \ .
\label{OP2}
\end{equation}
For the four column states $\Phi_{\bf i}$ has values 1,$-1$,1, and $-1$,
whereas it becomes zero for each of the plaquette-RVB states.
Consequently, the two crystalline bond states can be distinguished
by measuring two quantaties $\chi_\psi$ and $\chi_\phi$ simultaneously.
The order parameter (\ref{OP2}) coincides, in fact, with the magnetic
structure factor at the wave vector $(\pi,0)$ and has been studied to
check the difference between the two ordered magnetic phases:
$\Phi_{\bf i}\equiv 0$ in the N\'eel state, $\Phi_{\bf i}\neq 0$ 
in the collinear state. Thus, numerical results
for both $\chi_\psi$ and $\chi_\phi$ are currently available.
We refer to the most reliable data obtained on the $6\times6$
cluster \cite{exact2}. Fig.~3 of this work plots two
correlation functions 
$\chi_\psi(J_2)$ and $\chi_\phi(J_2)$. It is clearly
seen that at the peak point of $\chi_\psi(J_2)$ the other
function differs only slightly from its value at $J_2=0$,
whereas  $\chi_\phi(J_2)$ starts to increase simultaneously
with decreasing of $\chi_\psi$. This observation is a direct
convincing evidence of the plaquette-RVB ground state for the model (\ref{H}) 
at intermediate frustrations. 

We now turn to physical consequences of the translational symmetry
breaking in a magnetically disordered ground state of
spin system. The first property we address is a nature of
the phase transitions at $T=0$. If transformation from
such a disordered state to a spin ordered state, e.g., to the N\'eel
state, is continuous, the usual group-subgroup relations
should be satisfied for symmetry groups of the two phases.
However, this is not the case for the N\'eel and
the plaquette-RVB (or columnar dimer) states, since neither
the former has a higher symmetry than the latter (because 
of the absence of rotational invariance in spin space), nor
the latter is more symmetric than the former (due to 
translational symmetry breaking). Analogous arguments are applied
to the transition into the collinear state. Thus,
the magnetically disordered ground state region 
of the model (\ref{H})
is bounded by two points of first order transitions.
This conclusion explains, in particular, diverse estimates
of its width \cite{MFSWT,Schwb,Gochev,series,slbs} found by studying
{\it stabilities} of the different phases. 

Finally, we comment on finite temperature
behavior of valence bond crystals. As they break only the
discrete lattice symmetries, there should be transition from
the symmetric paramagnetic phase with decreasing temperature.
Analogous conclusion for the 
collinear state has been reached in
\cite{Ising}. The difference between these two cases lies 
in symmetry properties of the corresponding order parameters.
At $T\neq 0$, when sublattice magnetization of
the collinear phase vanishes, the order parameter is
a soft Ising type quantity $\Phi_{\bf i}$ having values
$+1$ and $-1$ on each lattice site. In this case only the
rotational lattice symmetry is broken in the low-temperature phase.
Singlet formation in the bond crystalline phases is accompanied
by translational symmetry breaking and corresponds to 
${\bf k}\neq0$. Bond strength
modulations appear at wave vectors $(\pi,0)$ or $(0,\pi)$
for the columnar dimerization and the spin bond
order parameter has two components:
${\bf S}_{\bf i} \cdot ({\bf S}_{{\bf i}-\hat{\bf x}}- 
{\bf S}_{{\bf i}+\hat{\bf x}})$ and 
${\bf S}_{\bf i} \cdot ({\bf S}_{{\bf i}-\hat{\bf y}}- 
{\bf S}_{{\bf i}+\hat{\bf y}})$.
For each of the four column phases only one component
is nonzero. 
The plaquette-RVB state 
also corresponds to these wave-vectors having
both components 
nonzero at the same time.
One can straightforwardly construct a two-component Landau
free-energy functional for this irreducible representation and show that
the choice between $(1,0)$ and $(1,1)$ symmetries
is determined by minimization of the `anisotropic'
fourth order term. Our calculations
for the energies of these two states at $T=0$ suggest that 
the minimum occurs for $(\pm 1,\pm 1)$ states. 

We thank O. A. Starykh and M. Troyer for helpful discussions.
This work has been financially supported by a Grant-in-Aid
from the Ministry of Education, Science and Cultutre of Japan.

\begin{figure}
\caption{ (a) plaquette-RVB and (b) columnar dimer ground states;
bold lines denote stronger spin bonds.}
\end{figure}

\begin{figure}
\caption{ Energies (in units of $J_1$) of
(a) columnar dimerized and (b) plaquette-RVB phases calculated
by bond-operator technique; (c) and (d) for the same phases by the
second-order perturbation theory.}
\end{figure}

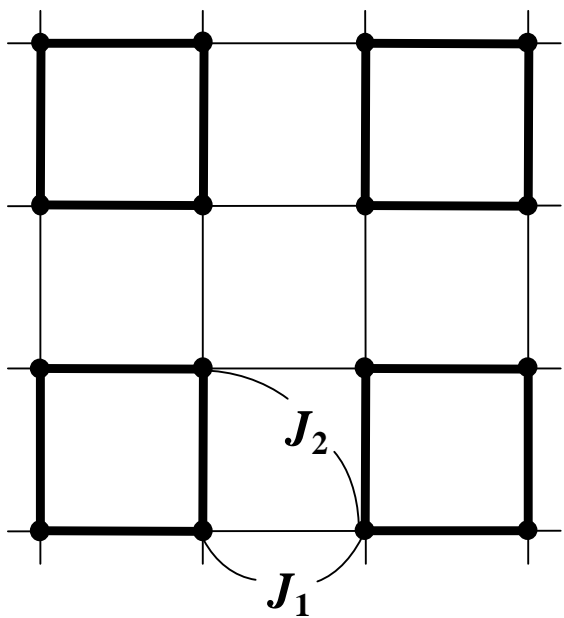
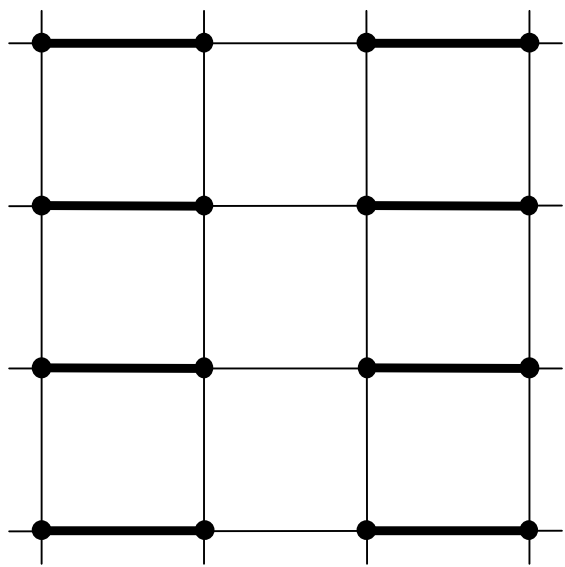
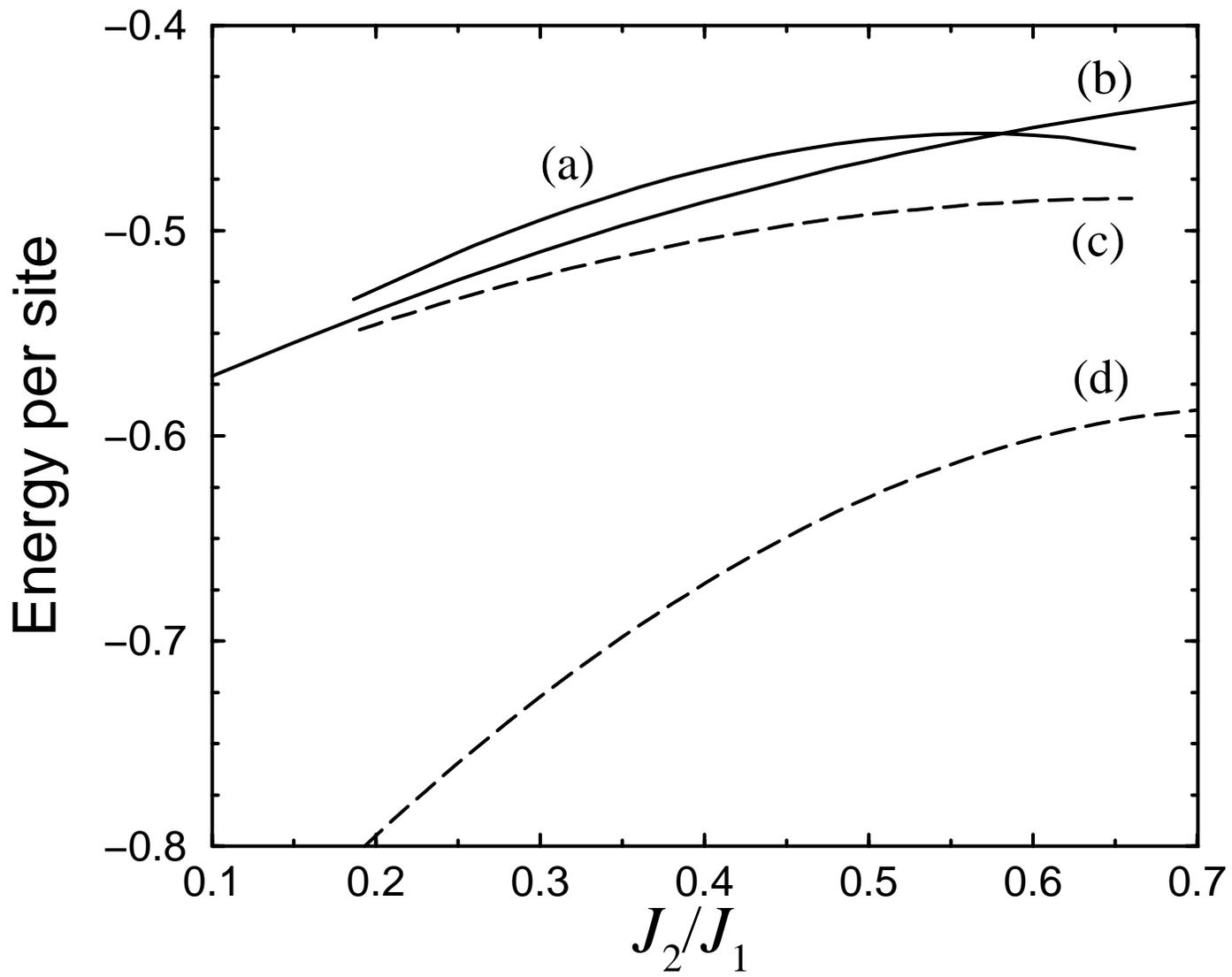
\end{document}